\documentstyle[twoside,fleqn,espcrc2,epsf]{article}
\def\fig#1#2#3{\epsfxsize=#3truein
\centerline{\epsffile{fig_#1.eps}}
\vskip 0.05 truein
\centerline{\vbox{{\bf \noindent Figure #1.} #2}}
\smallskip}

\def\figsizeA{2.3}
\def\figsizeB{2.6}
\def\figsizeC{2.1}

\def\cbc{\langle\overline{\chi}\chi\rangle}
\def\qbar{\overline{q}}
\def\qbq{\langle\overline{q}q\rangle}

\def\meff{m_{\rm eff}}
\def\mres{m_{\rm res}}

\def\MeV{{\rm\  MeV}}
\def\GeV{{\rm\  GeV}}

\def\spose#1{\hbox to 0pt{#1\hss}}
\def\ltapprox{\mathrel{\spose{\lower 3pt\hbox{$\mathchar"218$}}
 \raise 2.0pt\hbox{$\mathchar"13C$}}}
\def\gtapprox{\mathrel{\spose{\lower 3pt\hbox{$\mathchar"218$}}
 \raise 2.0pt\hbox{$\mathchar"13E$}}}
\def\inapprox{\mathrel{\spose{\lower 3pt\hbox{$\mathchar"218$}}
 \raise 2.0pt\hbox{$\mathchar"232$}}}

\def\one{$\qbq$ in units of $m_\gamma$ vs. $L_s$
from a numerical simulation of the dynamical two flavor 
Schwinger model with $m_0=0.9$, $m_f=0.0$ \cite{PMV_Schwinger}.
}
\def\two{Quenched QCD; $\mres^\lambda$ vs. $m_0$ from
an $8^4$, $\beta = 6.0$ configuration.
From top to bottom $L_s=8$, $L_s=12$ and $L_s=16$ 
\cite{Gadiyak_Ji_Jung}.
}
\def\three{Quenched QCD; $m_\pi^2(m_f \to 0)$ vs. $L_s$ with 
the plaquette action at $\beta = 5.7$ (filled)
and the Iwasaki action (open) at $\beta=2.2827$. 
\cite{Wu_latt99,RBC_quenched}.
}
\def\four{Quenched QCD; $\mres^R$ vs. $L_s$
with the plaquette action at $\beta = 5.65$ 
($a^{-1} \approx 1 \GeV$)
and $\beta = 6.0$ ($a^{-1} \approx 2 \GeV$) 
\cite{CP_PACS_chiral,Nagai_latt00}.
}
\def\five{Quenched QCD; $\mres^R$ vs. $L_s$
with the Iwasaki action at $\beta = 2.2$ ($a^{-1} \approx 1 \GeV$)
and $\beta = 2.6$ ($a^{-1} \approx 2 \GeV$) 
\cite{CP_PACS_chiral,Nagai_latt00}.
}
\def\six{Dynamical QCD; $\mres^G$ vs. $L_s$ with the
plaquette action on $8^3 \times 4$ lattices with  
$\beta=5.2$, $m_f = 0.02$, and $m_0 = 1.9$ (diamonds).
The cross/square are $\mres^G$/$\mres^\pi$ for
$8^3 \times 32$, $\beta = 5.325$ ($N_t=4$ crossover point)
\cite{Fleming_latt99,Columbia_thermo1}.
  }
\def\seven{Quenched QCD; $m_\pi^2$ vs. $m_f$ with the
plaquette action on $16^3 \times 32$ lattices with  
$\beta=5.7$, $L_s = 48$, and $m_0 = 1.65$ 
\cite{RBC_quenched,Wingate_latt00}. The star indicates
the $- \mres^R$ point. 
}
\def\eight{Quenched QCD; $-\qbq$ vs. $m_f$ with the
plaquette action on $8^3 \times 32$ (circles)
and $16^3 \times 32$ (squares) lattices with  
$\beta=5.7$, $L_s = 32$, and $m_0 = 1.65$ 
\cite{RBC_quenched,Mawhinney_latt00}.
}
\def\nine{Quenched QCD; $R = m_f  \chi_\pi / \qbq$ vs. $m_f$, where
$\chi_\pi$ is the chiral susceptibility. $R$ should be
unity in the absence of chiral symmetry breaking.
The bottom points are from unimproved DWF while the top
are from improved DWF by projecting out the 20 
slowest decaying eigenvectors \cite{Edwards_Heller_proj,Heller_latt00}. 
}
\def\ten{Quenched QCD; Non perturbative renormalization factors
vs. $(a p)^2$ on a $16^3 \times 32$ lattice with $\beta = 6.0$,
$L_s=16$, $m_0 = 1.8$. The top points correspond to $Z_{LL} / Z_A^2$.
The lower points correspond to factors that are zero
when chiral symmetry is exact \cite{Dawson_latt99,Dawson_latt00}.
}
\def\eleven{Quenched QCD; Baryon masses vs. $m_f$ on
a $16^3 \times 32$ lattice with $\beta = 6.0$, $L_s=16$
and $m_0=1.8$ \cite{Sasaki,Ohta_latt00}.
}
\def\twoelve{Quenched QCD; $Q_6^{(0)}$ vs. $m_f$ 
with the Iwasaki action on $16^3 \times 24$ lattices with  
$\beta=2.6$, $L_s=16$, and $m_0 = 1.8$ 
\cite{CP_PACS_chiral,Noaki_latt00}.
}
\def\thirteen{Dynamical QCD with 3 degenerate
flavors; $\qbq$ vs. the trajectory number
on a $16^3 \times 4$ lattice at the crossover point
$\beta=5.225$ with $L_s=32$, $m_f=0.02$ and $m_0 = 1.9$.
The top data is from a disordered initial configuration
while the bottom is from an ordered one \cite{Columbia_thermo2}.
}
\def\fourteen{Free QCD; energy density $\epsilon$ on the lattice 
over the continuum Stefan-Boltzman energy density $\epsilon_{SB}$
vs. the number of sites along the time direction $N_t$.
The spatial dimensions have $4 N_t$ sites,
$m_f=0.0$, and $m_0=1.0$ \cite{Fleming_latt00}.
}
\def\fifteen{Dynamical ${\cal N} = 1$ $SU(2)$
super Yang-Mills; $\cbc$ vs. $L_s$ with $m_f=0$ and $m_0 = 1.9$.  The
diamonds are from an $8^4$, $\beta=2.3$ lattice while the crosses
from $4^4$, $\beta=2.1$ \cite{Fleming_Kogut_Vranas,Fleming_Vietnam}.
}
\newcommand{\AmS}{{\protect\the\textfont2
  A\kern-.1667em\lower.5ex\hbox{M}\kern-.125emS}}

\title{Domain wall fermions and applications}
 
\author{Pavlos M. Vranas \thanks{
I would like to thank the  Lattice 2000 organizers 
for their support and hospitality.
Supported in part by NSF grant \# NSF-PHY96-05199.
Current address: IBM T.J. Watson Research Center,
Route 134, Yorktown Heights, NY 10598, USA.
vranasp@us.ibm.com.
}
\address{Physics Dept., University of Illinois, Urbana IL 61801, USA}
}

\begin{document}

\begin{abstract}
Domain wall fermions provide a complimentary alternative to
traditional lattice fermion approaches. By introducing an extra
dimension, the amount of chiral symmetry present in the lattice theory
can be controlled in a linear way. This results in improved chiral
properties as well as robust topological zero modes.  A brief
introduction on the subject and a discussion of chiral properties and
applications, such as zero and finite temperature QCD, ${\cal N} = 1$
Super Yang-Mills, and four-fermion theories, is presented.

\end{abstract}

\maketitle

\section{Introduction}
\label{sec_introduction}

When a continuum theory is regularized by the lattice the original
number of fermion species is increased by $2^d$, where $d$ is the
dimension of space-time, with a net chirality of zero. This is the
well--known fermion doubling problem \cite{Nielsen_Ninomiya}.  For
vector theories like QCD, this problem has been traditionally treated
at the expense of exact chiral symmetry by using Wilson \cite{Wilson_fermions}
or Kogut-Susskind \cite{Kogut_Susskind} fermions. Chiral symmetry is
broken for any non-zero lattice spacing $a$, but is recovered
together with Lorentz symmetry as $a \to 0$ resulting in the correct
target theory in the continuum.  However, since any numerical
simulation is done at non-zero $a$, one must be able to simulate at
small enough $a$ so that the effect of the breaking does
not obscure the underlying physics. The problem is that the
computational cost of decreasing $a$ is large.  For example, in the
full theory, decreasing $a$ by a factor of $2$ requires a factor of
$2^{8-10}$ more computations.

In recent years an alternative lattice fermion method, domain wall
fermions (DWF), has been developed and used in several
applications. The method was introduced by D.B. Kaplan
\cite{Kaplan_1,Kaplan_2} and was further developed by Neuberger and
Narayanan \cite{NN1} and by Shamir and Furman
\cite{Shamir,Furman_Shamir}.  Domain wall fermions are defined by
extending space-time to five dimensions.  A non--zero five dimensional
mass $m_0$ is present. The size of the fifth dimension is $L_s$ and
free boundary conditions for the fermions are implemented.  As a
result the two chiral components of the Dirac fermion are separated
with one chirality bound exponentially on one wall and the other
on the opposite wall. For any lattice spacing the two chiralities mix
only by an amount that decreases exponentially as $L_s \to \infty$.
At $L_s = \infty$ chiral symmetry is exact even at non-zero $a$. In
this way the continuum $a \to 0$ limit has been separated from the
chiral $L_s \to \infty$ limit. Furthermore, the computing cost
increases only linearly with $L_s$. Therefore, this method provides a
new practical ``knob'' on the amount of chiral symmetry breaking due
to the lattice discretization.

In order to have additional linear control over the ``effective''
fermion mass $\meff$, a four-dimensional mass $m_f$ is introduced by
explicitly mixing the plus chirality from one wall with the minus from
the other.  The parameters $L_s$, $m_0$ and $m_f$ control the
effective fermion mass $\meff$. In free one flavor theory one has
\cite{PMV_Schwinger}:
\begin{equation}
\meff = m_0 (2 - m_0) [ m_f + (1 - m_0)^{L_s}], \,
0 \langle m_0 \langle 2 \,
\label{meff}
\end{equation}
The $(1 - m_0)^{L_s}$ is a mass induced from the non zero overlap of
the two chiral components. This residual mass becomes zero as $L_s \to
\infty$. If $2< m_0 <4$ on has four flavors and if $4< m_0 <6$ six
flavors. There is a symmetry around $m_0 = 5$. In the presence of
increasing interactions these ranges get renormalized and shrink in
size. Furthermore, the decay rate is not just the simple $- \ln(1 -
m_0)$ of eq. \ref{meff} but also depends on the gauge coupling in a
complicated way.

The five dimensional theory contains $L_s$ heavy species. For small
$L_s$ these would not affect the dynamics.  However, as $L_s$ is
increased they may introduce bulk effects.  These effects must be
subtracted out by dividing the five-dimensional fermion determinant by
the determinant of the same operator without domain walls, i.e. with
anti-periodic boundary condition along the fifth direction
\cite{NN1}. This can be accomplished by introducing Pauli-Villars type
fields as in \cite{Furman_Shamir} or with the modification
\cite{PMV_Schwinger}.  Four-dimensional fermion fields $\qbar, q$ used
in Green's functions are constructed from the $s=0$ plus chirality and
$s=L_s-1$ minus chirality 
components of the five dimensional fields \cite{Furman_Shamir}.
Gauge fields are introduced by treating the $s$ direction as an
internal flavor space \cite{NN1,Kaplan_2}. They are only defined in
four-dimensions, and they do not change along $s$. This construction
is used by today's numerical simulations and is referred to as domain
wall fermions.

Because the gauge fields are constant across the fifth direction $s$,
the corresponding transfer matrix $T$ is independent of
$s$. Therefore, the product of the transfer matrices along $s$ is
simply $T^{L_s}$.  For $L_s \to \infty$ this becomes a projection
operator to the ground state defined by the corresponding
four-dimensional Hamiltonian $\cal H$. It turns out that $\cal H$ is
related to the hermitian Wilson Dirac operator $\gamma_5 D_W(m=-m_0)$
where $D_W$ is the Wilson operator.  Then the fermion determinant
$\det D$ of the five dimensional theory contains a factor of the form
$\langle b | 0 \rangle$. The boundary state 
$| b \rangle$ is independent of the gauge field
and has fixed filling level while $| 0 \rangle$ 
is the ground state of $\cal H$ and its filling 
level depends on the gauge field.  This
construction is the overlap formalism \cite{NN1} and has a powerful
consequence. If the filling level of $| 0 \rangle$ 
is different from that of $| b \rangle$ 
their overlap is exactly zero.  The fermion determinant $\det D$
can have exact and robust zero modes \cite{NN1}! The corresponding
index is simply the difference of the filling levels and is an
integer.  It turns out that:
\begin{equation}
{\rm index} = (K_{+} - K_{-}) / 2
\label{overlap_index}
\end{equation}
where $K_\pm$ is the number of positive/negative eigenvalues
of $\cal H$. For a finite lattice, $\cal H$ is a finite discrete
matrix amenable to numerical analysis.

Even though on the lattice there is no topology in the strict sense it
has been found that for smooth gauge field configurations the index
theorem is obeyed exactly \cite{NN1} and that for configurations
generated by numerical simulations of a quantum theory the index
theorem is obeyed in a statistical sense
\cite{Narayanan_Vranas,EHN_su3_index}. Also, it has been found
that the zero mode properties are maintained to a good degree for
relatively small $L_s$ \cite{DWF_zero_modes_Columbia}.

Unfortunately, the same property that makes DWF so attractive is also
the reason for a weakness. When a configuration changes from one
sector to another the number of negative eigenvalues of ${\cal H}$
changes and therefore an eigenvalue must cross zero. As a result $T$
develops a unit eigenvalue. In that case there is no decay along $s$,
the two walls do not decouple even as $L_s \to \infty$ and the chiral
symmetry can not be restored.  The set of configurations for which
$\cal H$ has a zero eigenvalue is of measure zero and therefore this
is not a problem \cite{NN1,Furman_Shamir}.  However, configurations in
their vicinity have very slow decoupling and as a result large values
of $L_s$ may be needed making the numerical simulations
expensive. This is a problem for large lattice spacings 
( for $a^{-1} \approx 650 \MeV$ an $L_s \approx 100$ may be needed) 
but it diminishes rapidly as $a$ becomes smaller
(for $a^{-1} \approx 2 \GeV$ an $L_s \approx 20$ 
is adequate for most problems).

The subject of these proceedings is to report on work done during the
past year relating to the chiral properties of DWF as well as to their
applications in QCD and other vector-like theories; for reviews from
previous years please see \cite{DWF_reviews} 

\section{How many ways?}
\label{sec_how_many}

Before continuing it
is important to put DWF into perspective. During the last few years a
wealth of new lattice fermions with improved chiral properties have
also been found.  One could now wonder; How many different methods are
there that solve the doubling problem for vector theories? Here is a
list:

\noindent
{\bf 1)} Wilson fermions, 1975 \cite{Wilson_fermions}.

\noindent
{\bf 2)} Kogut--Susskind fermions, 1975 \cite{Kogut_Susskind}.

\noindent
{\bf 3)} Domain wall fermions, 1992-1994 
\cite{Kaplan_1,NN1,Kaplan_2,Shamir,Furman_Shamir}.

\noindent
{\bf 4)} Infinitely many fields, 1992, 1998 \cite{Frolov_Slavnov,Slavnov}.

\noindent
{\bf 5)} Overlap fermions, 1992-1994 \cite{NN1}

\noindent
{\bf 6)} Neuberger fermions, 1997 \cite{Neuberger_fermions}.

\noindent
{\bf 7)} Perfect action fermions, 1997 \cite{P_Hasenfratz_GW}.

\noindent
{\bf 8)} Ginsparg-Wilson fermions, 1982, 1997 \cite{Ginsparg_Wilson,Luscher_GW}.

\noindent
{\bf 9)} Molecule chains, 1999 \cite{Creutz_molecules}.

\noindent
{\bf 10)} Topological QFT in 5D, 2000 \cite{Baulieu_Grassi_Zwanziger}.

These methods are substantially different in appearance but they are
of course related since they all describe the same continuum limit.
Furthermore, even at finite lattice spacing many of these methods are
intimately connected, some with exact analytical relations. Methods
3-10 share a common characteristic: in some limit of their parameters
an ``infinity'' is present allowing exact chiral symmetry even at
non-zero $a$. 

One may think that such a proliferation of methods is unnecessary.
However, further discoveries or improvements may come by exploiting
these different ways of solving the same problem. The following is a
quote from R. Feynman's ``The Character of Physical Law''
\cite{Feynman_quote}: ``Therefore psychologically we must keep all the
theories in our heads, and every theoretical physicist who is any good
knows six or seven different theoretical representations for exactly
the same physics. He knows that they are all equivalent, and that
nobody is ever going to be able to decide which one is right at that
level, but he keeps them in his head, hoping that they will give him
different ideas for guessing.''

A list of references for works presented in this conference on new
lattice fermions other than DWF is given in
\cite{Wenger_latt00}-\cite{Giusti_Hoelbling_Rebbi}.  The reader should
be aware that a large body of work on new lattice fermion methods is
available.

\section{Chiral properties}
\label{sec_chiral}

The chiral properties of DWF are determined by how well the chiral
modes are localized on the walls. How fast the modes decay away from
the wall is a dynamical property that depends on the other parameters
of the theory such as couplings, masses, domain wall height, and
volume, as well as the type of pure gauge action that is used. The
broad properties of this dependence have been known for some time.
However, work that was done during the past year has brought these
properties into much sharper focus
\cite{Nagai_latt00}-\cite{Vranas_Kogut_Tziligakis}.

Before proceeding, it is interesting to see how the decay rate depends
on the coupling constant, or equivalently the lattice spacing, in a
simple case: the two flavor Schwinger model.  In that model the chiral
condensate $\qbq$ must be exactly zero if chiral symmetry is exact. A
non zero value is a direct measure of breaking induced by the fermion
regulator. In figure 1 $\qbq$ in units of the photon mass $m_\gamma$
is plotted versus $L_s$.  The data are from a dynamical numerical
simulation \cite{PMV_Schwinger}. From top to bottom they correspond
to decreasing lattice spacings of $\sim 1/6$, $\sim 1/8$, $\sim 1/10$,
and $\sim 1/12$. The physical volume is kept fixed.
There are a few observations that can be made:

\vskip -0.3 in
\fig{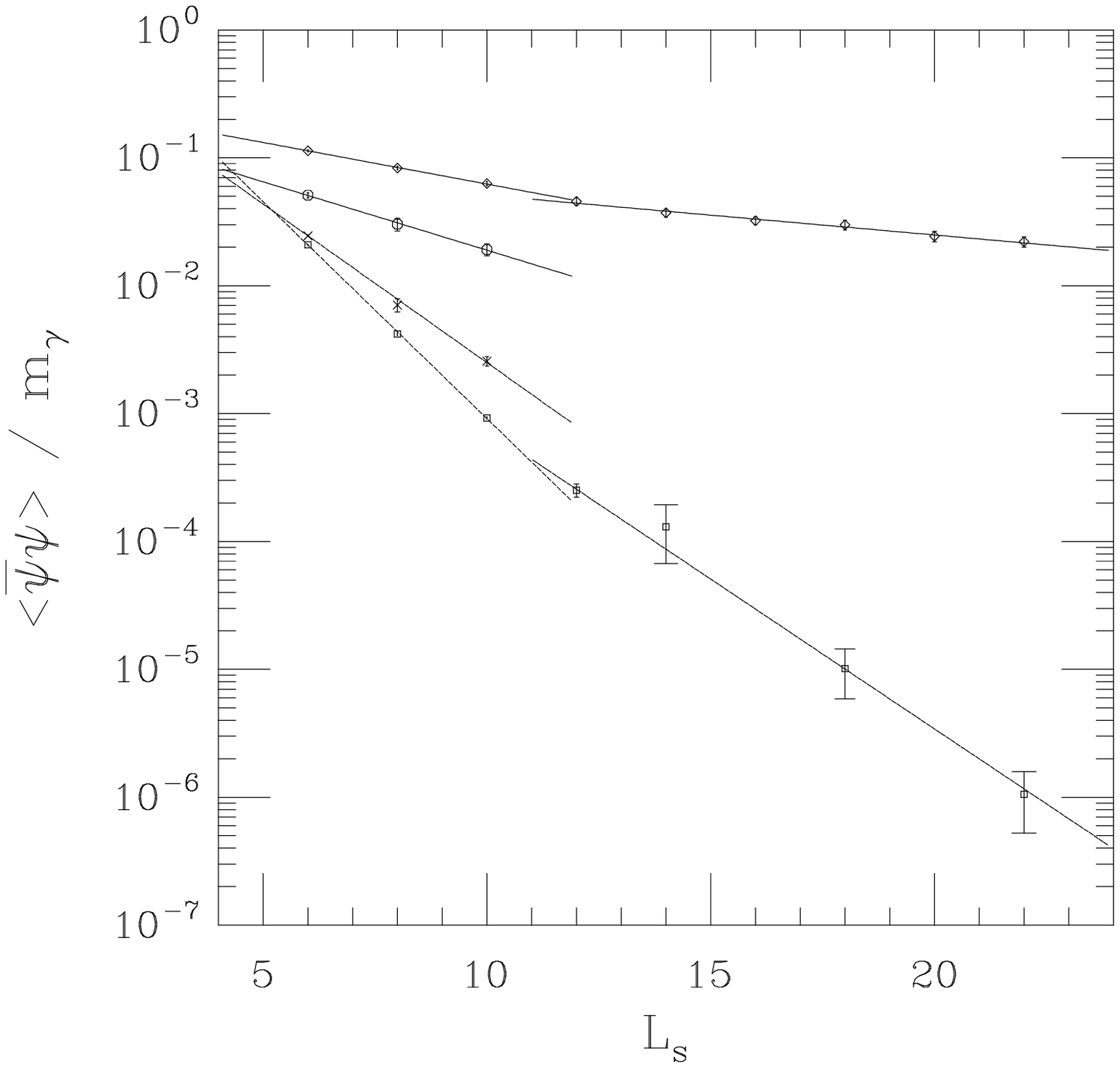}{\one}{\figsizeB}

\noindent
{\bf a)} The data is consistent with exponential restoration of chiral
symmetry.

\noindent
{\bf b)} There is a fast decay rate up-to $L_s \approx 10$, followed by a
slower one for larger $L_s$. The analysis in \cite{PMV_Schwinger}
suggested that the fast decay rate is controlled by the fluctuations
of the gauge field within a given topological sector, while the slower
decay is associated with topology changing configurations.
However, this also reflects the author's bias and does not exclude
a variety of other possible decay functions.

\noindent
{\bf c)} Both decay rates become faster as the lattice spacing is
decreased. This is important since otherwise numerical simulations
with DWF would not be very attractive.

Probing the effects of finite $L_s$ in QCD is much more challenging
because of spontaneous symmetry breaking.  Much of the progress during
the last year is due to inventive new probes that measure the residual
mixing between the two walls.  In a low energy effective Lagrangian
sense one expects that this mixing will be reflected in the presence of 
an additive residual mass term $\mres$ such that:
\begin{equation}
\meff = m_f +\mres(S_G, \beta, L_s, m_0, m_f, V)
\label{meff_int}
\end{equation}
where $S_G$ is the form of the pure gauge action, $\beta = 6 / g^2$ with $g$
the coupling constant and $V$ is the four-dimensional volume. The various 
approaches are:

\noindent
{\bf i)} From the pion mass:
\begin{equation}
m_\pi^2 \sim m_f + \mres^\pi \Rightarrow m_\pi^2(m_f = -\mres^\pi) = 0
\label{meff_pi2}
\end{equation}

\noindent
{\bf ii)} From PCAC \cite{Furman_Shamir} and fourth ref. in \cite{DWF_reviews}
\begin{equation}
\mres^R = 
{\langle \sum_x J^a_{5q}(x,t \gg 1)  J^a_{5q}(0,0) \rangle
\over {\langle \sum_x J^a_{5}(x,t \gg 1)  J^a_{5}(0,0) \rangle} }
\label{meff_pcac}
\end{equation}
where $J^a_5$ is the pseudo-scalar density, and $J^a_{5q}$ is constructed from
fields at $L_s/2$.

\noindent
{\bf iii)}
Using the Gell-Mann-Oaks-Renner (GMOR) relation 
\cite{Furman_Shamir,Fleming_latt99,Columbia_thermo1}
one can fit $\qbq$ vs. $m_f$ to the form:
\begin{equation}
(m_f + \mres^G) \chi_\pi = \qbq - b_0
\label{meff_gmor}
\end{equation}
where $\chi_\pi$ is the pseudo-scalar susceptibility
and $b_0$ is a constant.

\noindent
{\bf iv)}
By fitting the eigenvalues $\Lambda$ of the 5D hermitian DWF
operator vs. $m_f$ to the form:
\begin{equation}
\Lambda^2 =  n^2 [ \lambda^2 + (m_f + \delta m)^2], 
\,\,\, \mres^\Lambda = \langle \delta m \rangle 
\label{meff_lambda}
\end{equation}
where $n$, $\lambda$ and $\delta m$ are fit parameters \cite{RBC_quenched}.

\noindent 
{\bf v)}
By measuring the lowest eigenvalue $\lambda_{\rm min}$ of the 5D
hermitian DWF operator only for configurations where the
index is non-zero \cite{Lagae_Sinclair,Gadiyak_Ji_Jung,Jung_Edwards_Ji_Gadiyak,Jung_latt00}.  
\begin{equation}
\mres^\lambda = \langle \lambda_{\rm min} \rangle
\label{meff_lambda0}
\end{equation}
The index is measured using the overlap definition and the eigenvalue
flow method \cite{NN1}. For a non-zero index $\lambda_{\rm min}$
should be exactly zero in the $L_s \to \infty$ limit; for finite $L_s$
the deviation from zero is a measure of chiral symmetry braking.

Using these probes the dependence of $\mres$ on the parameters of the 
theory were studied in \cite{Nagai_latt00}-\cite{Vranas_Kogut_Tziligakis}.
In the following, a collection of sample results from these references
is presented. The figures are originals from the corresponding papers.
When the symbols used in the figures are different from the ones in this
article the correspondence will be noted in the figure caption.

\vskip -0.2 truein
\fig{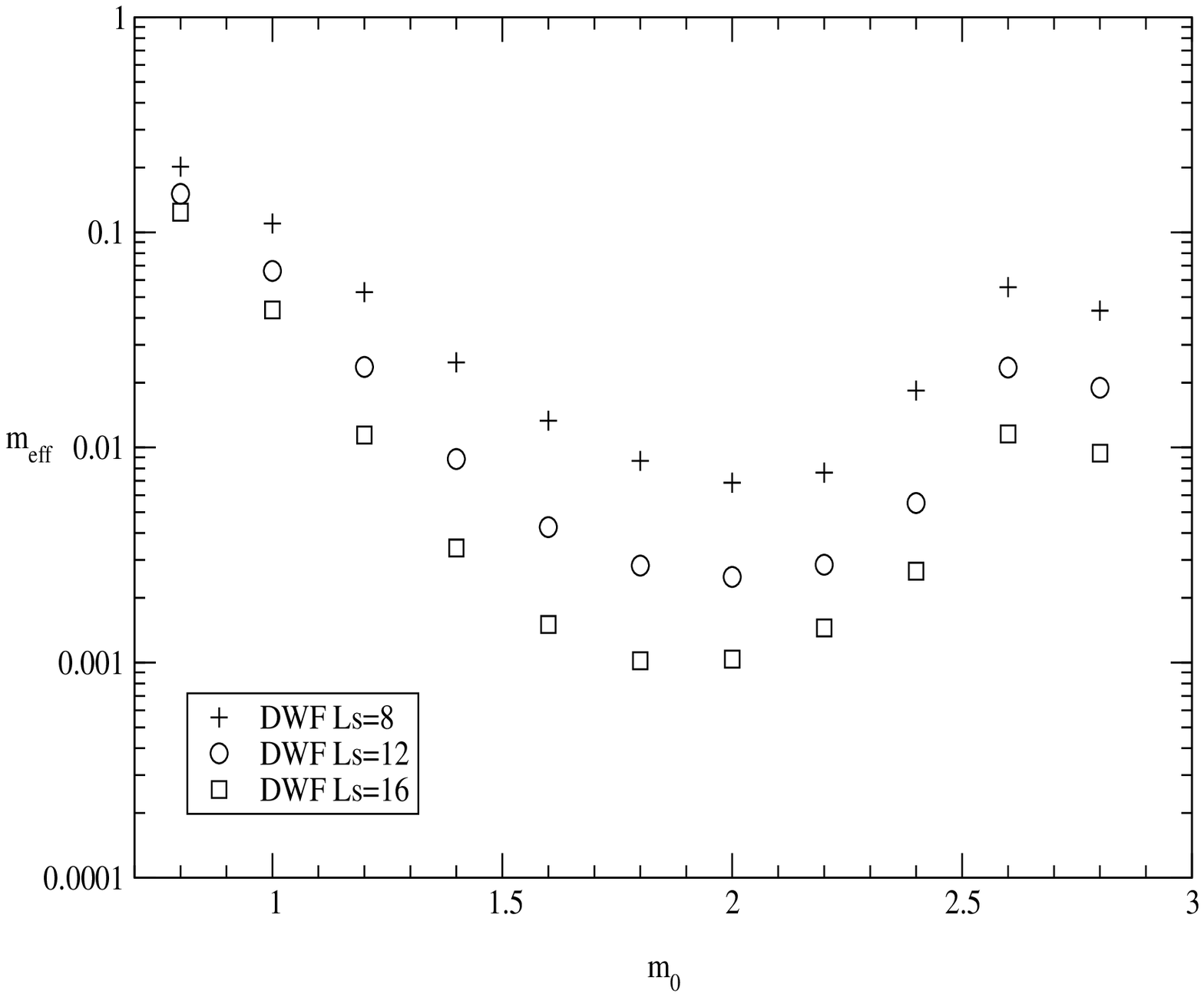}{\two}{\figsizeA}

The dependence of $\mres$ on $m_f$ is weak.  For example, this can be
clearly seen in  \cite{Aoki_Izubuchi_Kuramashi}.
The dependence on $m_0$ is shown, for example, in fig. 2
\cite{Gadiyak_Ji_Jung}. The value of $m_0$ is rather large except
in a plateau region between $1.5$ and $2.0$. Also, in that region
its value decreases faster with increasing $L_s$. The $L_s$ dependence
can be seen in fig. 3. The effect of chiral symmetry breaking decreases
with increasing $L_s$. Also, an Iwasaki improved gauge action
improves chiral symmetry substantially. The value of $m_\pi$ at $L_s=64$ with
the standard plaquette action is reproduced with the Iwasaki action 
with $L_s$ as small as $16$. This will be further discussed in
section \ref{sec_improvements}.

Perhaps the most important issue about DWF has to do with the way the
chiral limit is approached.  The overlap guarantees that the chiral
limit will be achieved as $L_s \to \infty$.  However, it is possible
that the allowed range of $m_0$ shrinks to zero size at strong
coupling. In \cite{Brower_Svetitsky} a strong coupling calculation
indicated that DWF lose their light state at very strong coupling.
Also, in \cite{Golterman_Shamir_2000} it was found that in the large
$m_0$ and strong coupling region DWF lose the light state.  At weaker
coupling the allowed range of $m_0$ is expected to be non-zero.  But
even then the precise way chiral symmetry is restored is not fully
understood. Obviously this is very important because one would like
to be able to fit the data using the full functional form.

\vskip -0.9 truein
\fig{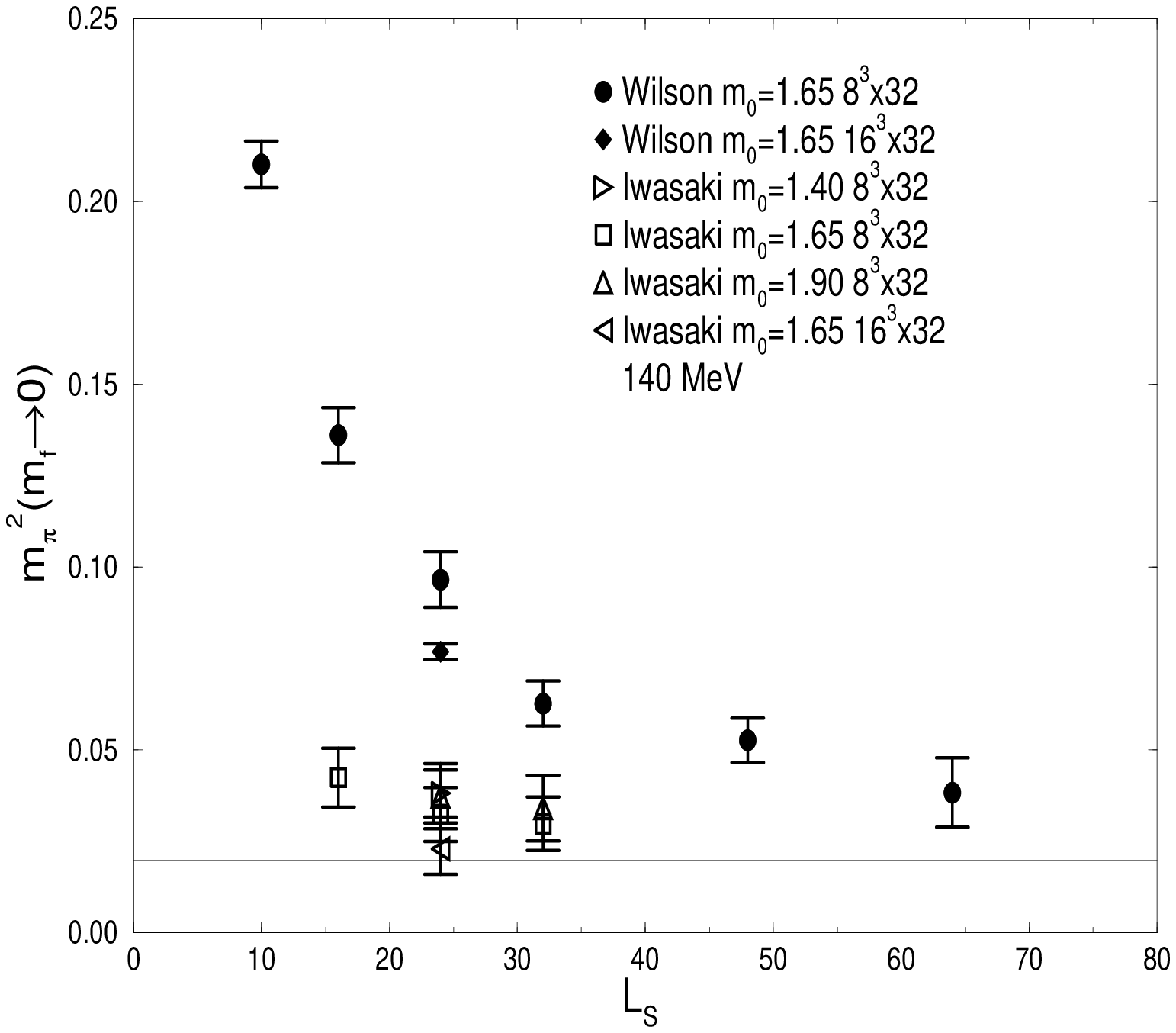}{\three}{\figsizeA}

The work of \cite{EHN_su3_index,EHN_decay} indicates that the the
spectrum of the 4-dimensional Hermitian Wilson operator has a non zero
density around zero. Although that density decreases fast with
decreasing lattice spacing it will induce some slow decay rates. This
was also found in \cite{Hernandez_Jansen_Luscher_note} where decay
rates as small as $10^{-3}$ where observed.  The work of
\cite{CP_PACS_chiral,Nagai_latt00} provides further insights: 
Results using the
Wilson plaquette action for $a^{-1} \approx 1 \GeV$ and $a^{-1}
\approx 2 \GeV$ are shown in fig. 4 while for the Iwasaki action and
for the same lattice spacings in fig. 5. Also, data points from
different volumes are shown indicating that the volume dependence is
weak.  Fits with exponential decay to zero were not good for $a^{-1}
\approx 1 \GeV$ Wilson, $a^{-1} \approx 2 \GeV$ Wilson and $a^{-1}
\approx 1 \GeV$ Iwasaki while fits with exponential decay to a
constant were reasonable.  For $a^{-1} \approx 2 \GeV$ Iwasaki an
exponential fit to zero was reasonable.

More data points may be needed in order to precisely establish the
form of chiral symmetry restoration as a function of $L_s$ for
different lattice spacings. It is possible that a more complicated
function of $L_s$ is required \cite{PMV_Schwinger,Shamir_new_act}.
For example, in \cite{RBC_quenched} several fitting functions were
used including double exponentials.

\vskip -0.2 truein
\fig{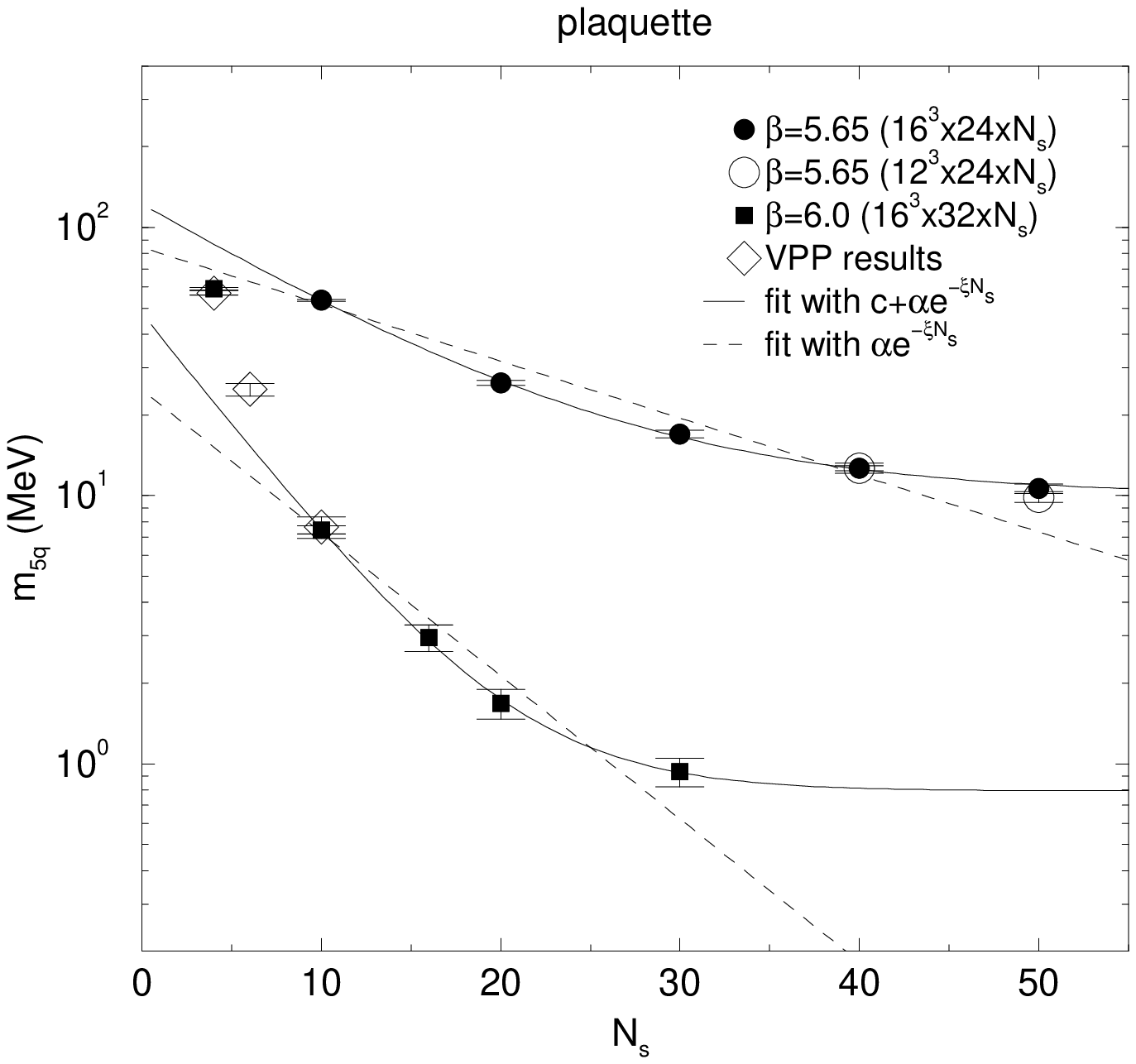}{\four}{\figsizeA}

\vskip -0.2 truein
\fig{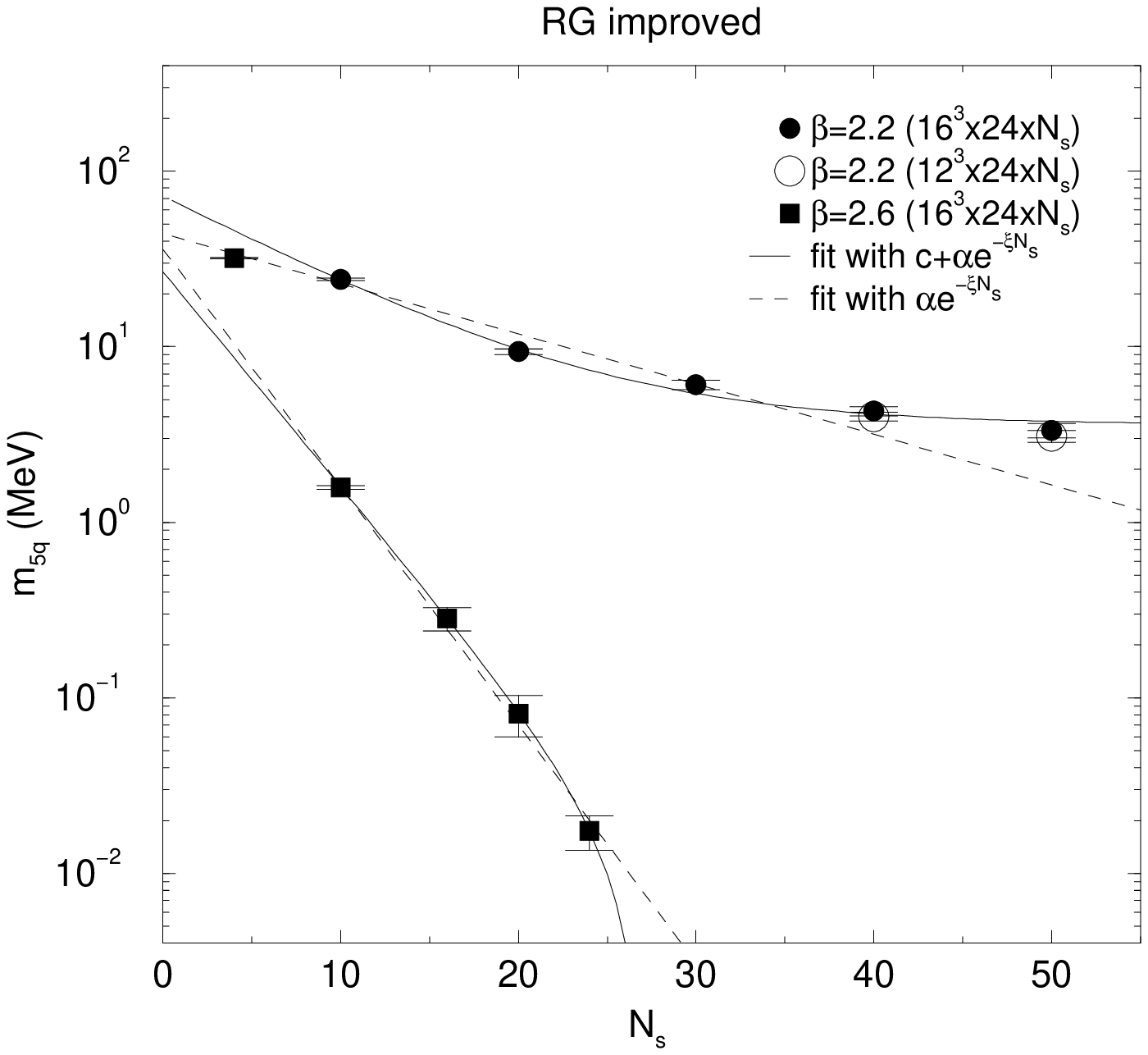}{\five}{\figsizeA}

The above QCD results are for the quenched theory. The
behavior of the dynamical theory has been studied in the
context of thermodynamics. Similar results are evident there
as can be seen from fig. 6.

The appeal of DWF is in the fact that any degree of chiral symmetry
can be achieved even at finite lattice spacing (provided that the
allowed $m_0$ region is not of zero size). However, in practice the
required value of $L_s$ may be large making the calculations
difficult. An alternative approach was introduced in
\cite{RBC_quenched}.  At finite $L_s$ it is possible to approach the
chiral limit by tuning $m_f$ so that $m_f + \mres = 0$. This approach
is very similar to Wilson fermions. Even the Aoki phase
is present for $m_f + \mres < 0$ 
\cite{Vranas_Kogut_Tziligakis,Izubuchi_Nagai,Aoki_Izubuchi_Kuramashi}. 
The same chiral symmetry
breaking terms are present but with much smaller coefficients.

\vskip -0.3 truein
\fig{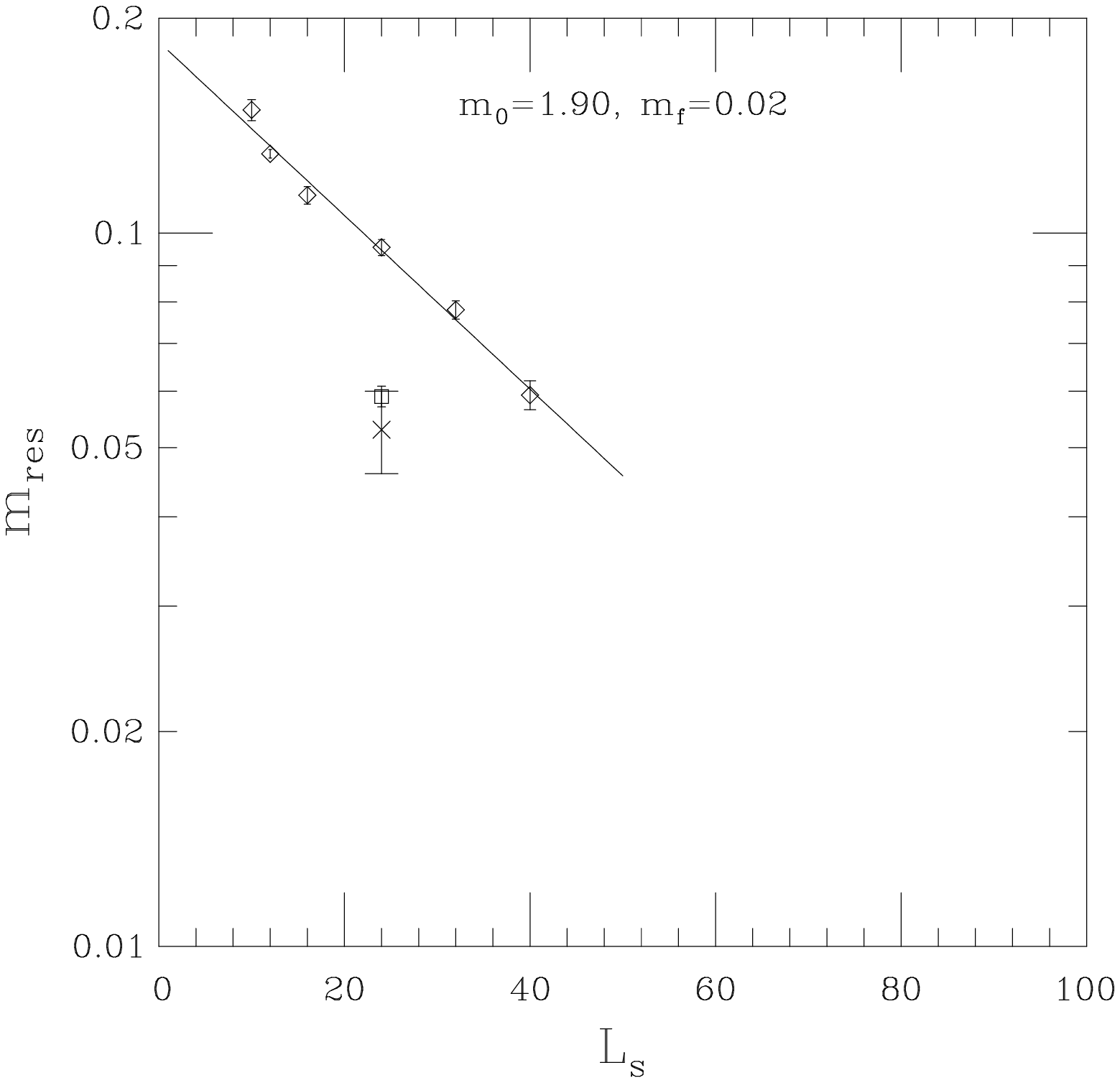}{\six}{\figsizeB}

\fig{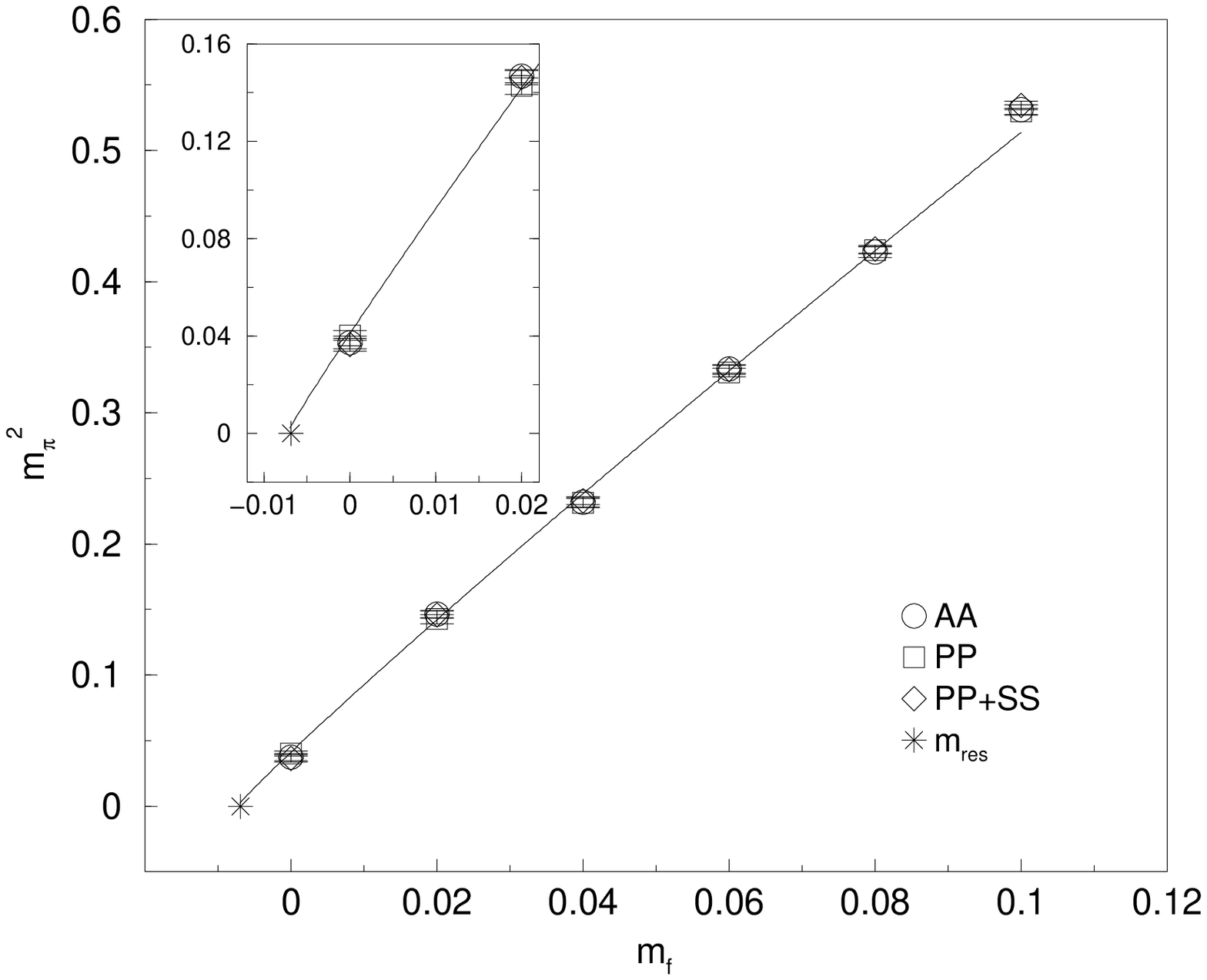}{\seven}{\figsizeA}

In fig. 7 \cite{RBC_quenched,Wingate_latt00} the approach of the pion mass to zero 
for quenched QCD is shown vs. $m_f$.
This approach may be distorted by: order $a^2$ effects,
finite $L_s$ effects in the ultraviolet, topological near-zero
modes (due to quenching), finite volume effects, and quenched
chiral logs. In the quenched theory
configurations with near zero modes are not suppressed by the fermion
determinant. Because of the improved chiral properties of DWF
the propagator has a near-pole as $m_f \to 0$ causing 
the pseudo-scalar pseudo-scalar (PP) correlator to diverge. 
This dramatic effect is shown in fig. 8 for $\qbq$ \cite{RBC_quenched}.
However, the coefficient of the divergent term
is expected to vanish like $1 / \sqrt{V}$ where $V$ is the volume.
This is evident in fig. 8 \cite{RBC_quenched,Mawhinney_latt00} 
(a similar divergence is present above the
quenched deconfining transition but it does
not disappear as $V$ is increased \cite{Columbia_quenched_thermo}).
Therefore, the pion mass extracted from the PP correlator for small
volumes may be polluted.  This can be alleviated by extracting the pion
mass from the axial axial (AA) correlator instead \cite{RBC_quenched}.

Another pathology of quenching is the presence of quenched
chiral logs. This can be accommodated by fitting to a function that
contains a log as in fig. 7. This fit gives $\mres^\pi = 0.0073(10)$
which is consistent with the value calculated using PCAC 
$\mres^R = 0.0072(9)$ \cite{RBC_quenched}. 
This provides a self-consistency check
for the understanding of chiral symmetry breaking with DWF.

\fig{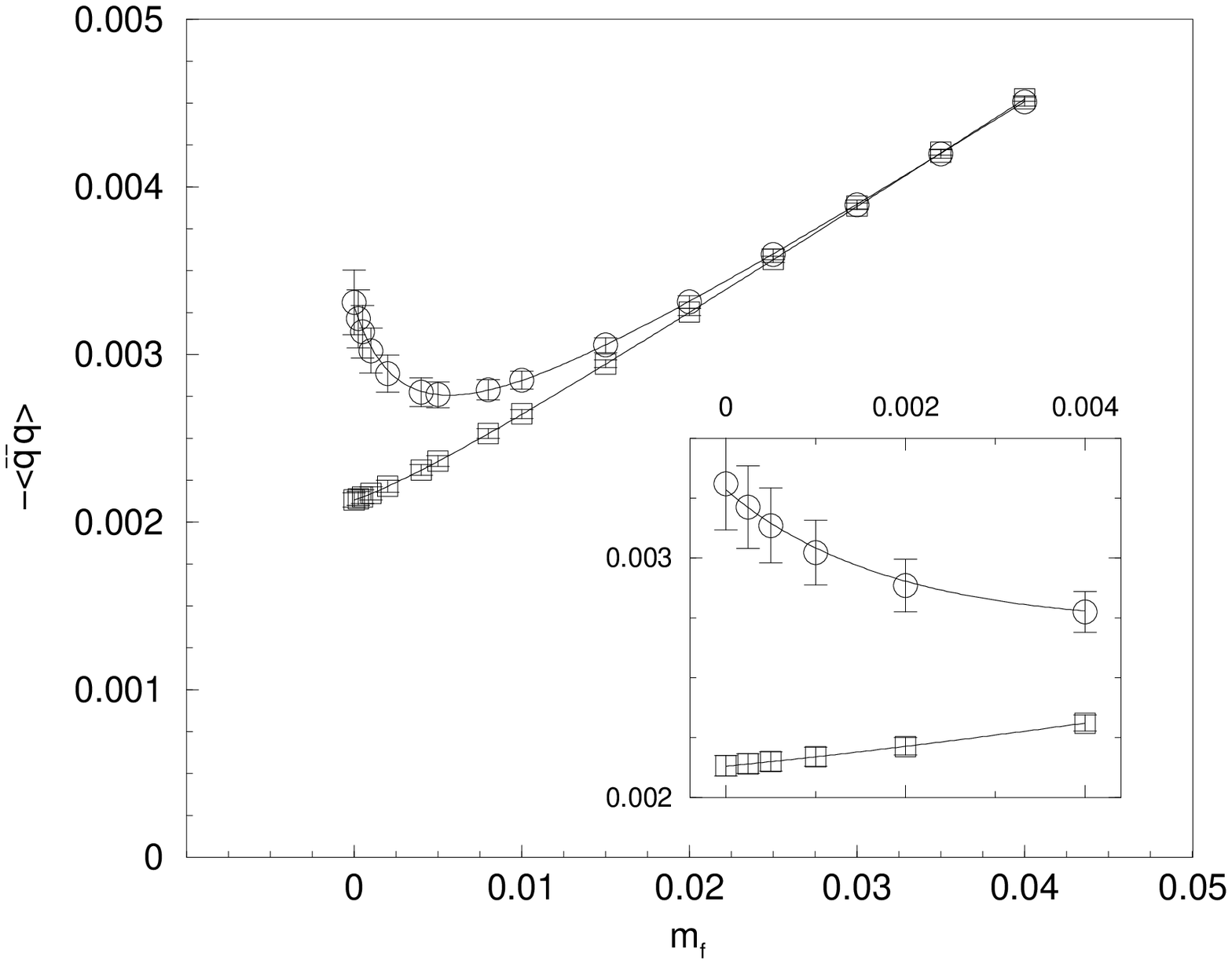}{\eight}{\figsizeA}

However, it must be stressed that $m_f + \mres = 0$ does not eliminate
all chiral symmetry breaking effects \cite{Mawhinney_latt00}. Power
divergent operators introduce $O(\mres/ a^2)$ effects that must be
subtracted out.  These effects can be clearly seen in
\cite{Mawhinney_latt00} where $f_\pi^2 m_\pi^2 / 48(mf+\mres)$ does
not extrapolate to $\qbq(m_f + \mres = 0)$.

\section{Improvements}
\label{sec_improvements}

In this section improvements to DWF are discussed.  The improved
methods achieve the same degree of chiral symmetry
restoration but require less computing. Several methods have
been developed during the last year 
\cite{Wu_latt99,Heller_latt00,Edwards_Heller_proj,Hernandez_Jansen_Luscher_note},
\cite{Borici}-\cite{Narayanan_Neuberger_5d}.
Also, work on locality properties during the past year can be found in
\cite{Hernandez_Jansen_Luscher_locality}-\cite{Kikukawa_bounds}.

As seen in section \ref{sec_chiral} the chiral properties of DWF are
significantly better when an Iwasaki pure gauge action is used.  This
is remarkable; a change in the pure gauge action results in an
improvement in the fermion sector.  This is probably because the
Iwasaki action ``over-improves'' and therefore reduces the frequency of
index changes due to small instantons shrinking below the lattice
spacing and disappearing. As mentioned in section
\ref{sec_introduction}, index changes result in slow decay rates. This
is indicated by the work of \cite{EHN_su3_index} where it was found that
index changes at values of $m_0$ where simulations are done are mainly
due to small objects of size around the lattice spacing.

Using the Iwasaki action at $a^{-1} \approx 2 \GeV$ and $L_s=20$
reduces $\mres$ by a factor of $10$ \cite{CP_PACS_chiral}.  At $a^{-1}
\approx 1 \GeV$ and $L_s=20$ the improvement is not as dramatic,
$\mres$ is reduced by a factor of $2$ \cite{Wu_latt99,CP_PACS_chiral}.
At $a^{-1} \approx 650 \MeV$, which is the
lattice spacing around the $N_t=4$ thermal
transition, and $L_s=24$, $\mres$ reduces only by a
factor of $1.2$. Therefore, the Iwasaki action
is effective at smaller $a$.
QCD thermodynamics with $N_t=4$ does not benefit much from it but
at $N_t=6$ ($a^{-1} \approx  1 \GeV$)
one would expect a factor of two improvement.
Similar results have been obtained with other improved gauge
actions \cite{EHN_su3_index,Jung_Edwards_Ji_Gadiyak}.

If these interpretations are correct one would expect that actions
used for ``cooling'' studies may be beneficial. For example, one could
attempt to reduce index changes by ``over-improving'' the Iwasaki
action even more. However, a study at $a^{-1} \approx 650 \MeV$ with
an Iwasaki coefficient $c=-1$ instead of $c=-0.331$ did not show any
significant further improvements \cite{Christian_Vranas}.  Also, a
study at $a^{-1} \approx 650 \MeV$ with an action that restricts the
plaquette to $1 - {\rm Tr}[U_p] / 3 < 0.8$ did not show significant
further improvements either \cite{Fleming_Vranas}.  Nevertheless, 
these actions at smaller lattice spacings may produce
better results.

Another set of improvements has to do directly with the fermion
sector:

\noindent
{\bf 1)} The projection method
\cite{Edwards_Heller_proj,Narayanan_Neuberger_5d,Hernandez_Jansen_Luscher_note}
consists of projecting out a few eigenvectors with the slowest decay
rate.  This method works remarkably well as can be seen from fig. 9.

\noindent
{\bf 2)} An overlap inspired method introduces a different 5-dimensional
operator \cite{Narayanan_Neuberger_5d}. Although the computational
requirement is similar to DWF this method is better for analytical
work.

\noindent
{\bf 3)} A perturbation theory inspired method uses beyond nearest
4D neighbor action \cite{Shamir_new_act}. It
has better decay rates at weaker couplings.

\noindent
{\bf 4)} An algorithm that takes advantage of the absence
of interactions along $s$ uses multi-grid methods
to speed-up computations \cite{Borici}.

\fig{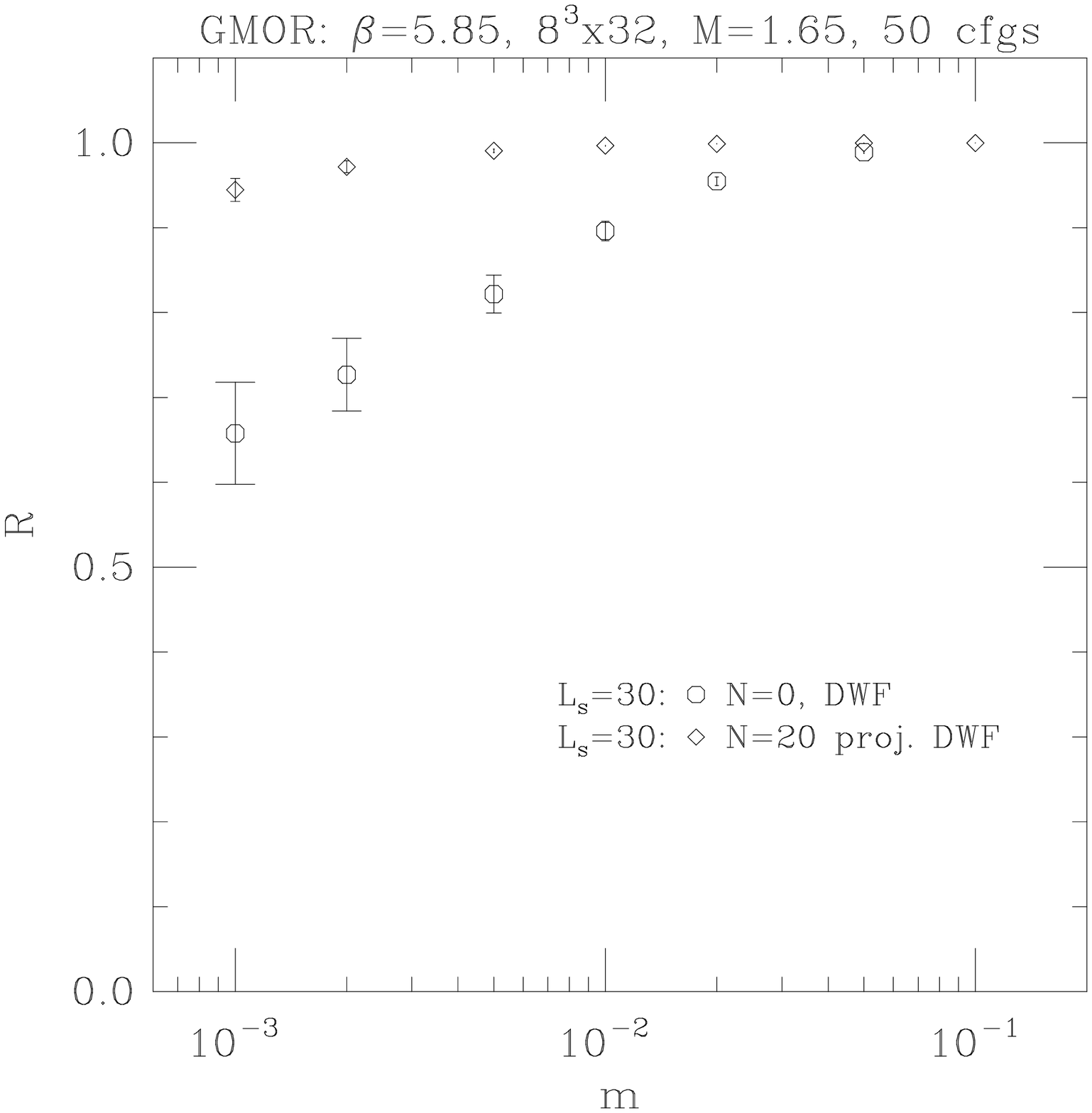}{\nine}{\figsizeC}

\section{Quenched QCD}
\label{sec_quenched}

Several applications of DWF to quenched QCD were done during the past
year \cite{Wingate_latt00}-\cite{Ichinose}. Most of them have been
presented in the corresponding reviews but they are also mentioned
here for completeness.

Calculations of $f_{\pi, K}$ have been made in
\cite{Aoki_Izubuchi_Kuramashi,RBC_quenched} (see also the review
\cite{Lellouch_latt00}).  For example, in \cite{RBC_quenched} it was found that
$f_{\pi, K}$ are close to the experimental values. Two different
methods were used to calculate $f_\pi$; one used the PP correlator and
the value of $\mres^R$ while the other used the AA correlator and did
not require a value for $\mres$. It was found that $(f_\pi)_{PP}
/(f_\pi)_{AA} = 1.00(10)$. This provides a strong consistency check
for the determination of $\mres$.

A calculation of the nucleon to rho mass ratio \cite{RBC_quenched}
gave $m_N / m_\rho = 1.37(5)$
in the $a \to 0$ limit and for a box with size $\approx 1.6$ fermi
in each direction (the experimental value is $1.22$). Also, the chiral
condensate was measured to be  $\qbq = (245(7) \MeV)^3$ at $L_s = 16$
and $\qbq = (256(8) \MeV)^3$ at $L_s = 24$ (the phenomenological estimate is
$(229(9) \MeV)^3$).

\vskip -0.3 truein
\fig{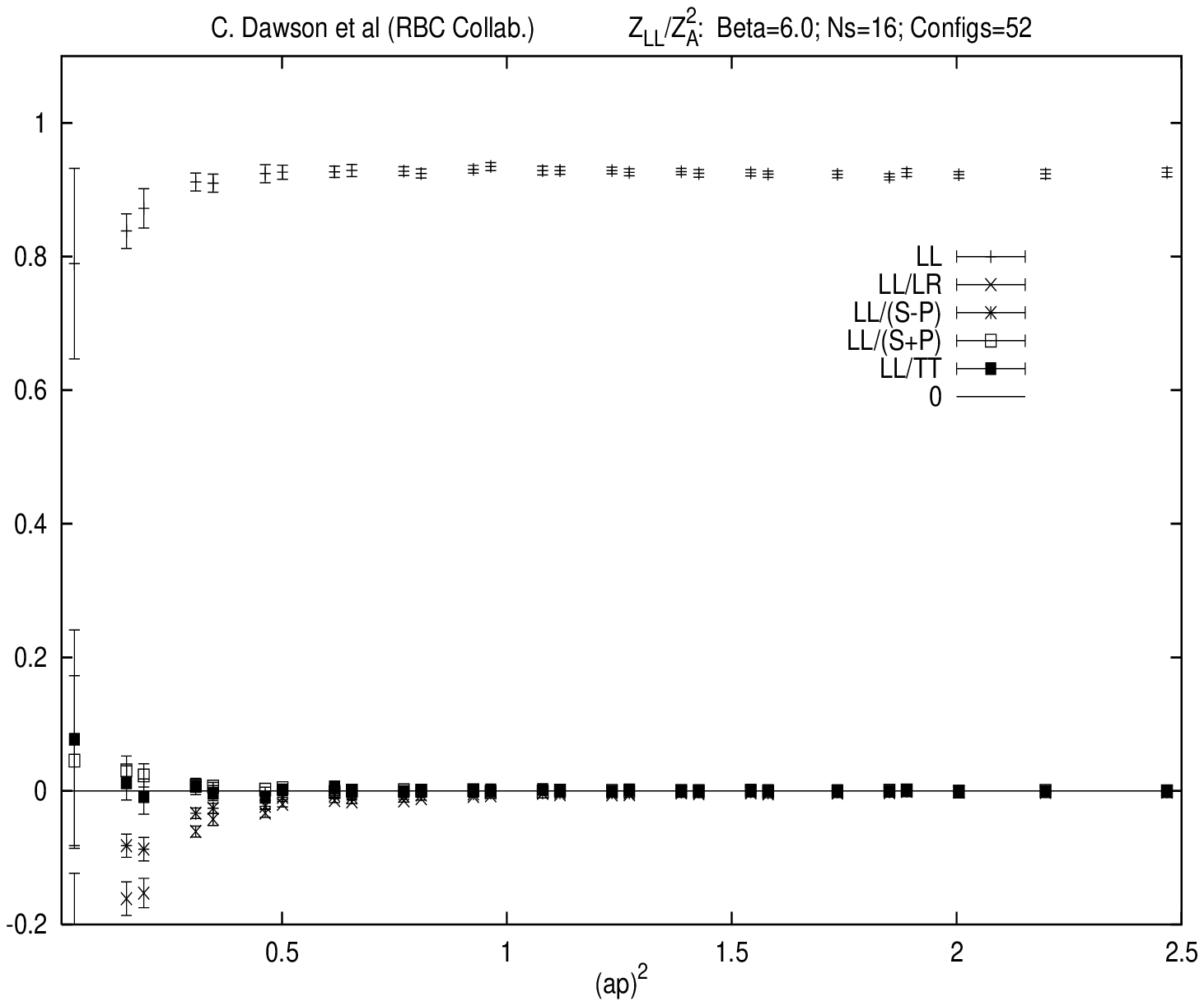}{\ten}{\figsizeA}

\fig{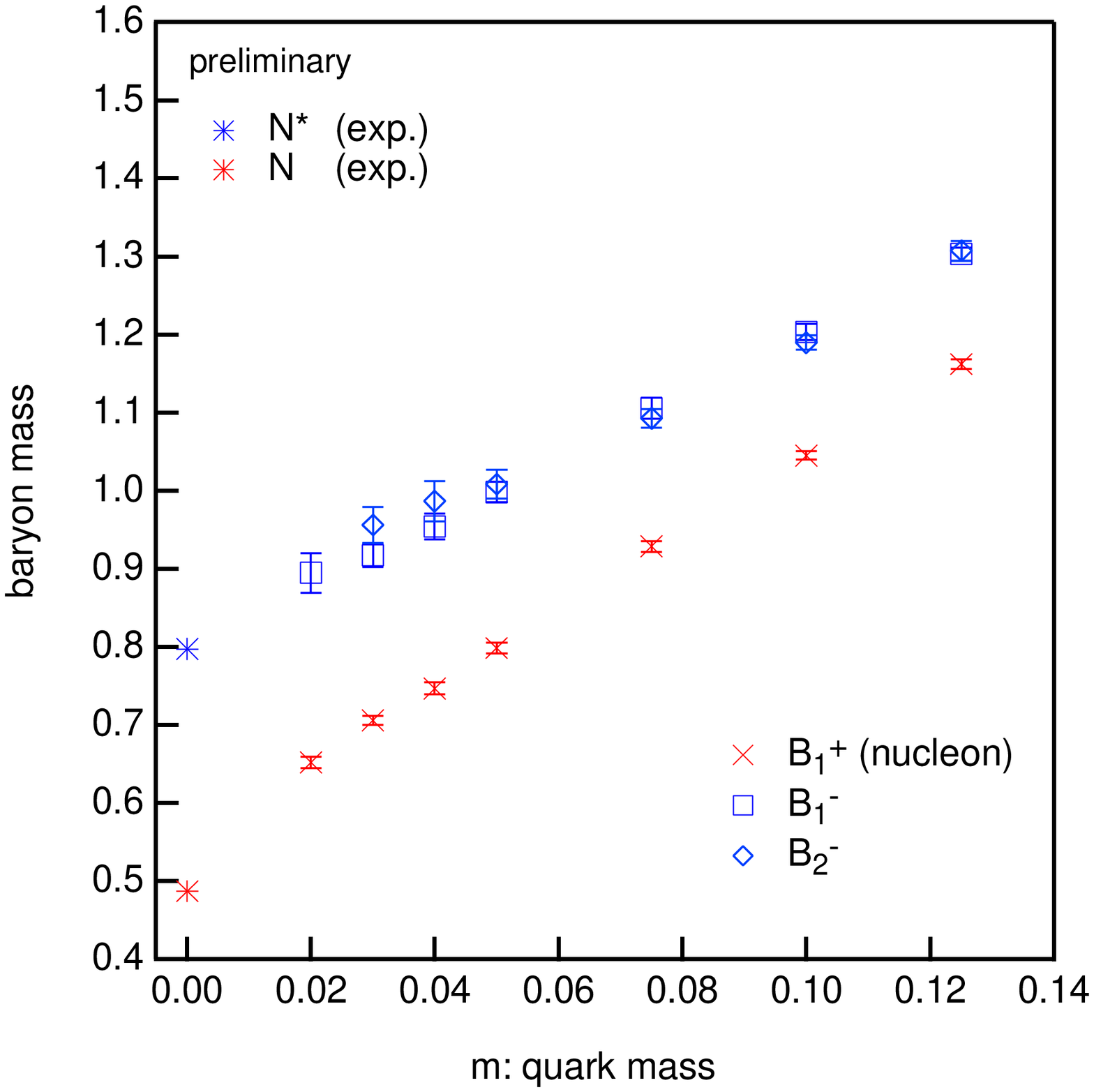}{\eleven}{\figsizeA}

The renormalization factors relevant to weak matrix elements were
calculated using non-perturbative methods in
\cite{Dawson_latt99,Dawson_latt00} (see also the review
\cite{Sint_latt00}).  In fig. 10 one can see that the $Z_{LL} / Z_A^2$
factor is not zero as expected. The other factors in that figure are
required to vanish if chiral symmetry was exact.  Indeed, they are
consistent with zero demonstrating the good chiral properties of DWF.
Perturbative calculations of these factors were done in
\cite{Aoki_Izubuchi_Kuramashi_pert_ren,Aoki_Izubuchi_Kuramashi}.
Also, a method that extends the range over which operators relevant to
weak matrix elements are defined was developed in
\cite{Zhestkov_latt00} (also see the review \cite{Sint_latt00}).

The mass splitting between the nucleon $N(939)$ and its parity partner
$N^*(1535)$ was calculated in \cite{Sasaki,Ohta_latt00} 
and is shown in fig. 11.
The data are consistent with the experimental values.

\fig{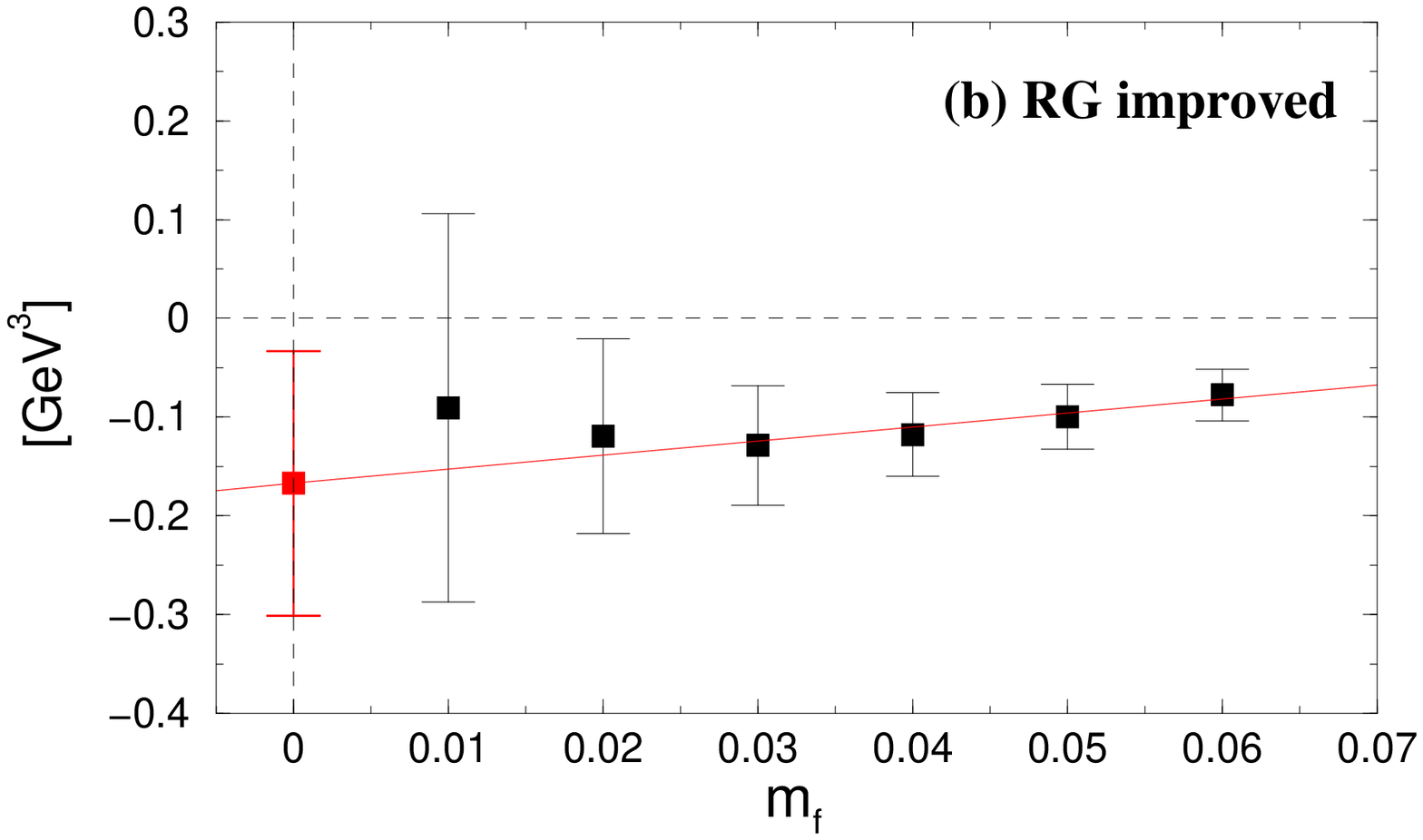}{\twoelve}{\figsizeA}

The value of $B_K$ was computed non-perturbatively in the RI scheme
and matched to the $\overline{MS}$-$NDR$ scheme in
\cite{Blum_latt00}. It was found to be $B_K(2 \GeV) = 0.538(8)$ on a
$16^3 \times 32$ lattice with $L_s = 16$. This is lower than the value
$B_K = 0.562(14)$ obtained by \cite{Taniguchi_latt00}, probably
because the non-perturbative renormalization constant of
\cite{Blum_latt00} is lower than the perturbative one used in
\cite{Taniguchi_latt00} (see also the review \cite{Lellouch_latt00}).

Operators relevant to $\epsilon\prime / \epsilon$ have also been
calculated
\cite{Noaki_latt00,Blum_latt00,Mawhinney_latt00,Dawson_latt00} (see
also the review \cite{Lellouch_latt00}).  As an example the
$Q_6^{(0)}$ operator is shown in fig. 12 \cite{Noaki_latt00}. The
magnitude of this operator is small.

\section{Dynamical QCD thermodynamics}
\label{sec_thermo}

The study of the QCD thermal transition can obviously benefit a great
deal by using lattice fermions with better chiral properties.  DWF are
a natural candidate for such studies and have been used over the past
few years to study $N_t = 4$ thermodynamics.  This effort crystallized
during the past year \cite{Columbia_thermo1,Vranas_Dubna}, 
\cite{Fleming_latt00}-\cite{Vranas_latt98}. The main findings are:

\noindent 
{\bf 1)} A small but non-zero $U(1)_A$ symmetry breaking
above and close to the transition is present.
The good zero-mode properties of DWF lent validity to this result.

\noindent 
{\bf 2)} Two degenerate flavor simulations on $16^3 \times 4$ lattices with
$L_s = 24$ gave a transition temperature of
$T_c = 163(4) \MeV$ and $m_\pi = 427(11) \MeV$, where the $\rho$ mass
was used to set the scale.  Simulations with ordered and disordered
starting configurations agreed after thermalization indicating the
absence of a first order transition for this value of $m_\pi$.

\noindent 
{\bf 3)} Three degenerate flavor simulations on $16^3 \times 4$ lattices
with $L_s = 24$ \cite{Columbia_thermo_Nt4_3f} gave a transition
temperature of $T_c = 160(3) \MeV$ and $m_\pi = 402(7) \MeV$.  Again,
simulations with ordered and disordered initial configurations
indicated the absence of a first order transition for this value of
$m_\pi$. The time history of these simulations at the crossover point
is shown in fig. 13.

\vskip -0.2 truein
\fig{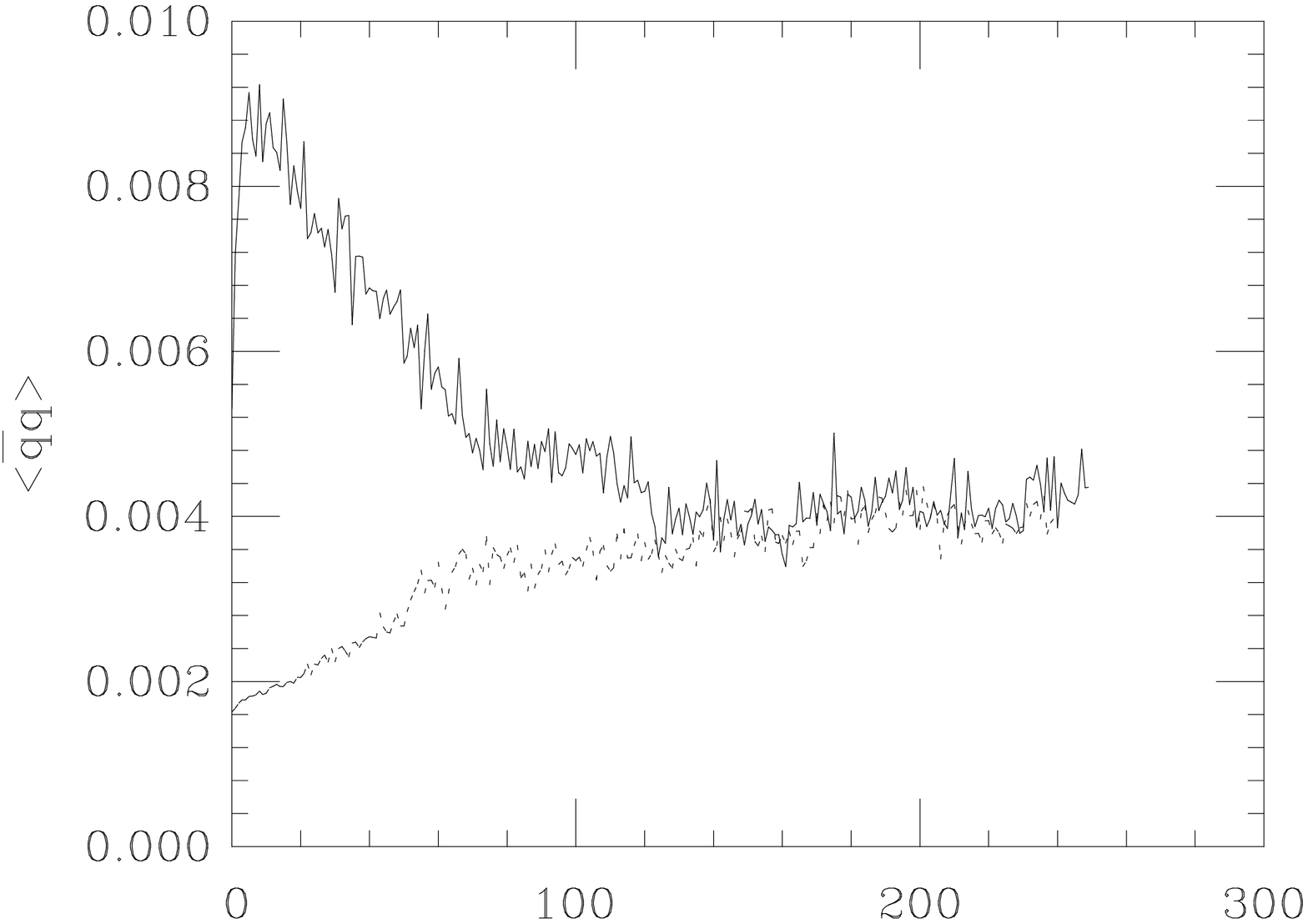}{\thirteen}{\figsizeC}

Clearly, $m_\pi$ is very large. From fig. 6 it was estimated that $L_s
\approx 100$ may be needed to make $m_\pi \approx 200 \MeV$. On the
other hand, the transition with $N_t = 6$ will be at $a^{-1} \approx 1
\GeV$.  From quenched studies at $a^{-1} \approx 1 \GeV$, see for
example fig. 3, one would expect that the use of an Iwasaki action
would be beneficial and that one could achieve $m_\pi \approx 200
\MeV$ with $L_s = 24$.

A study of the equation of state using DWF could be less demanding.
One could integrate along the path proposed by G. Fleming
\cite{Fleming_latt00}.  Integrate from $\beta = 0$ to a $\beta$ above
the transition using the quenched theory which exactly corresponds to
$m_f = 1$ because of the subtraction of the heavy fields. Keeping
$\beta$ fixed at that value one can then integrate along a line of
decreasing $m_f$ using the dynamical theory. Since this is done 
at realtively small $a$ an $L_s \approx 24$ may be enough to make the
calculation physically relevant.

\vskip -1.0 truein
\fig{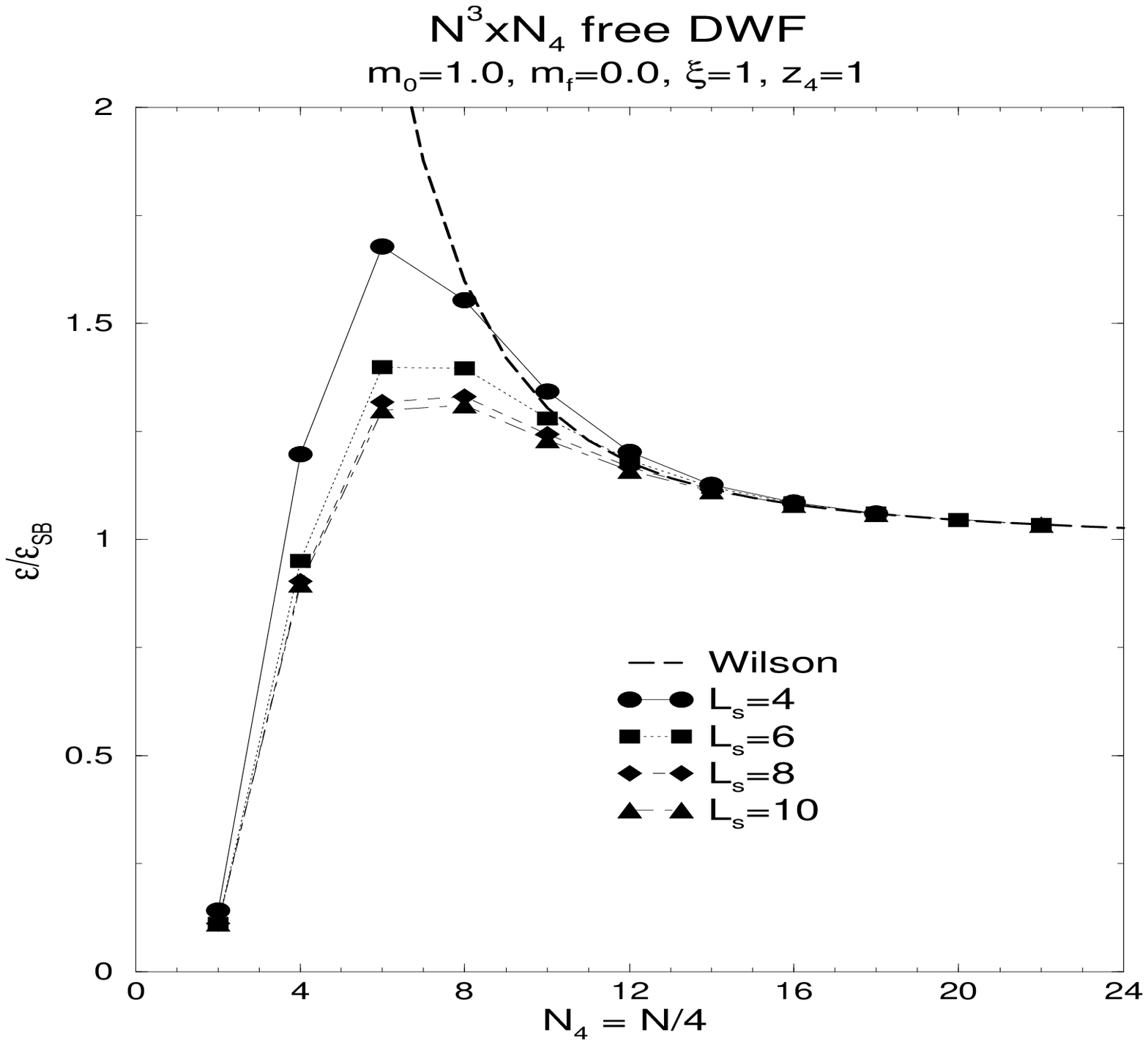}{\fourteen}{\figsizeC}

In order to understand how good DWF are with $N_t=4$ or even $N_t=6$
the free fermion energy density is shown in fig. 14
\cite{Fleming_latt00}. Due to the subtraction of the heavy modes DWF
perform better than Wilson fermions at small $N_t$.

\section{Super Yang-Mills}
\label{sec_sym}

The ${\cal N} = 1$ super-symmetric Yang-Mills (SYM) theory can be simulated
on the lattice using traditional techniques, very much like in QCD.
Simulations using Wilson fermions have already been performed, see for
example \cite{Wilson_SUSY}.  However, these methods as in QCD, require
fine tuning in order to recover the target theory in the continuum.

DWF offer an alternative \cite{Neuberger_fermions,Kaplan_Schmaltz} and
\cite{Nishimura}-\cite{Fleming_Vietnam}. 
At $m_f=0$, $L_s \to \infty$ DWF forbid a
gluino mass term and because all other symmetry breaking operators are
irrelevant SUSY is recovered in the continuum limit without
fine-tuning.  Furthermore, DWF for $m_f \ge 0$ have a Pfaffian (Pf)
with definite sign for any gauge field configuration. This is
important since in a numerical simulation the Pf is used as a
Boltzmann weight which must be of definite sign.  Wilson fermions at
finite lattice spacing do not have a Pf of definite sign and
numerical simulations are done with $|{\rm Pf}|$ instead. Since Pf in the
$a \to 0$ limit has a definite sign one expects to recover the target
theory but one may worry about non-analyticities resulting from taking
the magnitude.

\fig{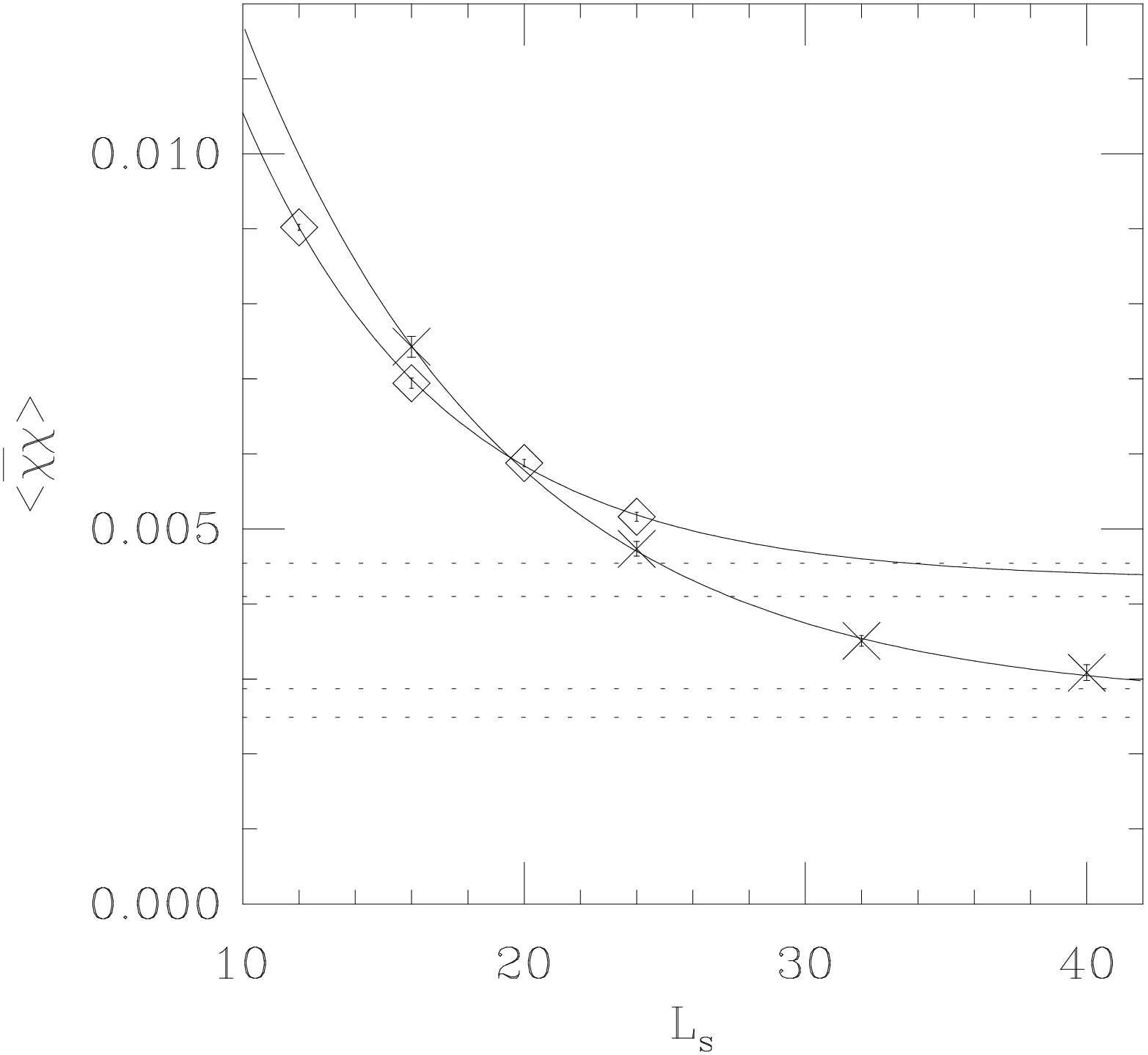}{\fifteen}{\figsizeC}

For ${\cal N}=1$ SYM the index is equal to $2 \nu N_c$ where $\nu$ is
the topological charge and $N_c$ the number of colors.  As a result,
instantons break the $U(1)$ chiral symmetry down to $Z_{2 N_c}$ and an
operator with $2 N_c$ gluino fields acquires a vacuum expectation
value. An interesting question is what happens to the remaining $Z_{2
N_c}$ symmetry. It is expected that $Z_{2 N_c}$ may break
spontaneously to $Z_2$ \cite{Witten}.  
However, on a torus there are
constant field solutions with fractional topological charge $1/ N_c$
\cite{tHooft_torons}.  Although for large volumes these solutions
vanish, for small volumes they may play a role and induce a VEV even
for zero mass.  In particular, for $m V \cbc \ll 1$ the $\nu = \pm 1 /
N_c$ sectors exclusively contribute to $\cbc \ne 0$ \cite{Leutwyler_Smilga}.

Numerical simulations of ${\cal N}=1$ SU(2) SYM were done in
\cite{Fleming_Kogut_Vranas}. The results for $m_f=0$ are shown in
fig. 15. As can be seen, $\cbc(L_s \to \infty)$ has a non zero VEV
which is maintained even for rather small volumes. The time histories of
the $4^4$ volume and $m_f=0$ had ``spikes'' which are indicative of
zero mode effects. These results seem to support a non-zero $\cbc$ due
to configurations with fractional topological charge.  Fractional
topological charge configurations have already been found in the
quenched theory \cite{EHN_fract}.

\section{Fermion--scalar interactions}
\label{sec_compare}

All DWF fields across the extra direction interact the same way with
the gauge field. 
The interaction of DWF with scalar fields was 
studied in \cite{Vranas_Kogut_Tziligakis} and was found to be
different. That interaction takes place
only along the link that connects the boundaries of the extra
direction. This reveals a richness in the way different spin particles
couple to DWF. Four-fermion
models were studied using large N techniques and were
supported by numerical simulations with N=2. It was found that the
chiral properties of DWF in these models are good
across a large range of couplings and that a phase with parity-flavor
broken symmetry can develop for negative $m_f$ if $L_s$ is finite.

\section{Conclusions}
\label{sec_conclusions}

For the first time domain wall fermions separate the
continuum ($a \to 0$) from the chiral ($L_s \to \infty$) limits. 
Since the computing requirement is only linear in $L_s$ they provide
practical control over chiral symmetry. Furthermore, they exhibit
robust zero modes which become exact at the $L_s \to \infty$
limit. 

DWF provide a complimentary alternative to traditional fermion methods
and can shed light to different regions of the parameter space. DWF
have found a large spectrum of applications such as: QCD
thermodynamics, quenched QCD, Super Yang-Mills, four-fermion theories,
and the Schwinger model. Also, there are proposals for improving
DWF to achieve the same amount of chiral symmetry with
less computations.

Finally, it should be noted that DWF are just one of the many new
lattice fermion methods. This wealth of approaches 
can only lead to further new discoveries.

\end{document}